\documentclass[aps,prb,reprint,showpacs,
preprintnumbers,bibnotes,superscriptaddress]{revtex4-2}
\usepackage{amsmath,amssymb,amsfonts}
\usepackage{bm}
\usepackage{ulem}
\usepackage{mathrsfs}
\usepackage{array,dcolumn}
\usepackage{booktabs}
\usepackage{longtable}
\usepackage{graphicx}
\usepackage{pgfplots}
\pgfplotsset{compat=1.16}
\usepackage{bm}
\usepackage{ulem}
\usepackage{chemformula}
\usepackage{hyperref}
\usepackage{txfonts}
\usepackage{lipsum}
\usepackage{appendix}

\usepackage{tabularx}
\usepackage{multirow}

\usepackage{xcolor}
\begin{document}

\title{Chern numbers of topological phonon band crossing determined with inelastic neutron scattering}

\author{Zhendong Jin}
\thanks{These authors contributed equally to this study.}
\author{Biaoyan Hu}
\thanks{These authors contributed equally to this study.}
\author{Yiran Liu}
\thanks{These authors contributed equally to this study.}
\affiliation{International Center for Quantum Materials, School of Physics, Peking University, Beijing 100871, China}

\author{Yangmu Li}
\affiliation{Beijing National Laboratory for Condensed Matter Physics, and Institute of Physics, Chinese Academy of Sciences, Beijing 100190, China}
\affiliation{Condensed Matter Physics and Materials Science Division, Brookhaven National Laboratory, Upton, New York 11973, USA}
\affiliation{University of Chinese Academy of Sciences, Beijing 100049, China}

\author{Tiantian Zhang}
\affiliation{Beijing National Laboratory for Condensed Matter Physics, and Institute of Physics, Chinese Academy of Sciences, Beijing 100190, China}

\author{Kazuki Iida}
\author{Kazuya Kamazawa}
\affiliation{Neutron Science and Technology Centre, Comprehensive Research Organisation for Science and Society (CROSS), Tokai, Ibaraki 319-1106, Japan}

\author{A. I. Kolesnikov}
\author{M. B. Stone}
\affiliation{Neutron Scattering Division, Oak Ridge National Laboratory, Oak Ridge, Tennessee 37831, USA}

\author{Xiangyu Zhang}
\author{Haiyang Chen}
\author{Yandong Wang}
\affiliation{State Key Laboratory for Advance Metals and Materials, University of Science and Technology Beijing, Beijing 10083, China}

\author{I. A. Zaliznyak}
\author{J. M. Tranquada}
\affiliation{Condensed Matter Physics and Materials Science Division, Brookhaven National Laboratory, Upton, New York 11973, USA}

\author{Chen Fang}
\email{cfang@iphy.ac.cn}
\affiliation{Beijing National Laboratory for Condensed Matter Physics, and Institute of Physics, Chinese Academy of Sciences, Beijing 100190, China}

\author{Yuan Li}
\email{yuan.li@pku.edu.cn}
\affiliation{International Center for Quantum Materials, School of Physics, Peking University, Beijing 100871, China}
\affiliation{Collaborative Innovation Center of Quantum Matter, Beijing 100871, China}

\date{\today}

\begin{abstract}
Topological invariants in the band structure, such as Chern numbers, are crucial for the classification of topological matters and dictate the occurrence of exotic properties, yet their direct spectroscopic determination has been largely limited to electronic bands. Here, we use inelastic neutron scattering in conjunction with \textit{ab initio} calculations to identify a variety of topological phonon band crossings in MnSi and CoSi single crystals. We find a distinct relation between the Chern numbers of a band-crossing node and the scattering intensity modulation in momentum space around the node. Given sufficiently high resolution, our method can be used to determine arbitrarily large Chern numbers of topological phonon band-crossing nodes.
\end{abstract}

\maketitle
\section{Introduction}
Ever since the discovery of topological quantum numbers in quantum Hall states \cite{Klitzing1980,Thouless1982}, the concept of band topology has shed light on the exploration and classification of crystalline materials \cite{Qixiaoliang2011,Burkov2016,Bansil2016,chiuchingkai2016,Armitage2018}. Topological insulators, semimetals, and superconductors are extensively studied, both as novel phases of matter and for their potential applications. Unlike conventional phases of matter described by symmetry in the Landau paradigm, topological phases are classified by topological invariants, which do not change over adiabatic deformations of the band structure.

An important topological invariant is called the Chern number, which is associated with a mapping from a two-dimensional (2D) closed surface in reciprocal space to the Hilbert space of Bloch states. The Chern number characterizes the topological structure of such mapping and has observable consequences. In the gapped energy spectrum of 2D quantum Hall systems, nonzero Chern numbers correspond to the number of edge states which lead to the quantization of the Hall conductance \cite{Andreas2009,Qixiaoliang2011}. In three-dimensional (3D) Weyl semimetals, Weyl nodes act as monopoles of Berry flux and have nonzero Chern numbers (defined by the mapping from their enclosing surface in momentum space to the Hilbert space), which dictate the number of Fermi-arc surface states \cite{Wan2011,Yang2015,Lv2015} and the quantized magnitude of circular photogalvanic effect \cite{deJuan2017,Flicker2018}. As band topology is independent of the statistics of the constituent quasiparticles, similar phenomena are also expected in bosonic systems. For example, topological photonic and acoustic bands and their corresponding surface states have been found in artificial structures \cite{Haldane2008,Lu2013,Lu622,Xiao2015,Ge2018}. In natural crystals, a variety of Dirac and Weyl nodes have been predicted and/or observed in phonon \cite{He2018,Li2018,Zhangtiantian2018,Miaohu2018,Lijiaxu2018,Xia2019,li2020} and magnon bands \cite{Mook2016,Likangkang2017,Yao2018,Bao2018,McClarty2022}. Previous experiments on these systems mainly focused on the bulk dispersion relation near the band-crossing points, rather than on topological invariants such as the Chern numbers, partly because it is difficult to measure phonon and magnon surface states and transport behaviors.

\begin{figure*}
\centering{\includegraphics[width=\textwidth]{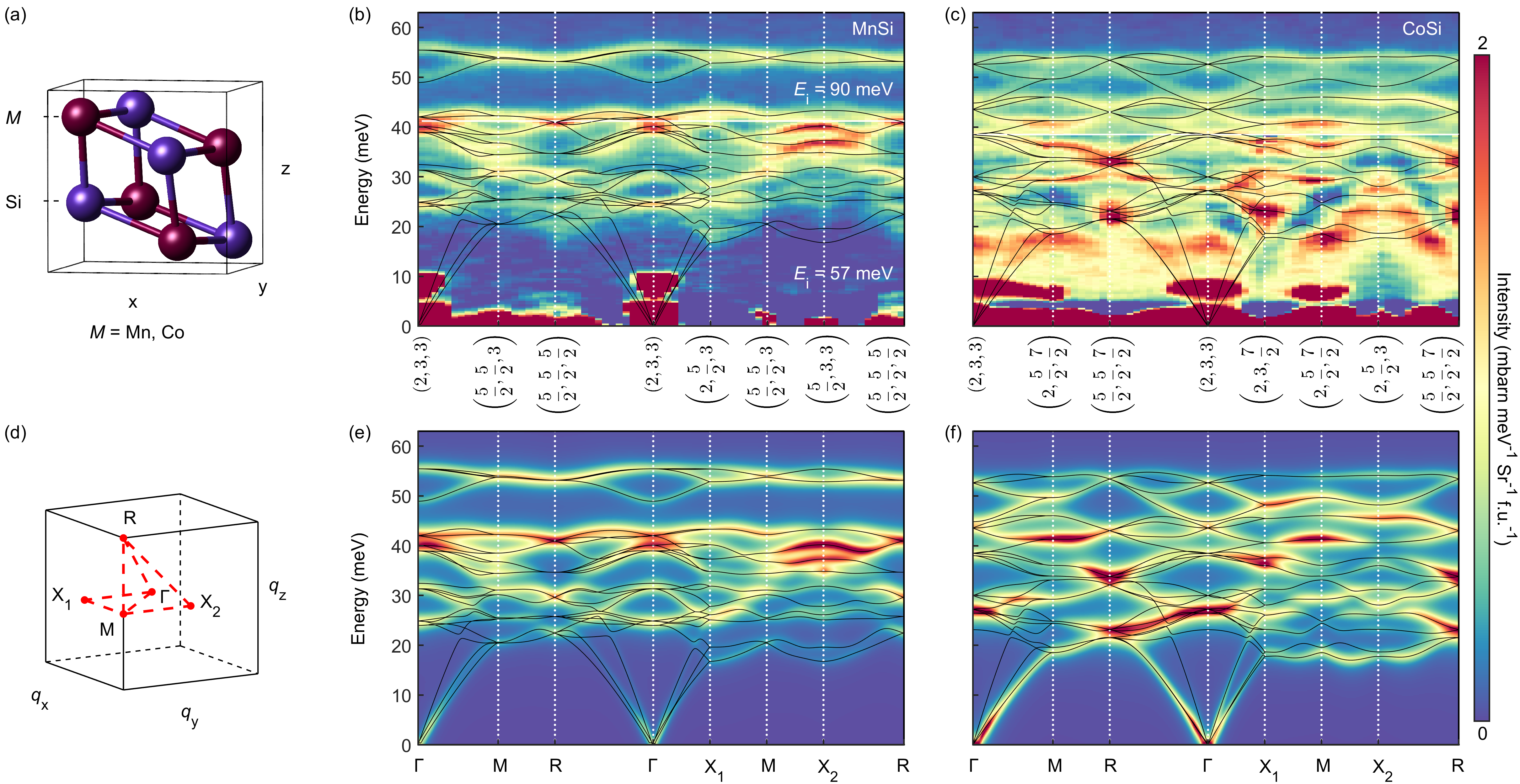}}
\caption{\label{fig1} (a) Cubic primitive cell of $M$Si ($M$ = Mn, Co). (b) and (c) Representative INS intensities of MnSi and CoSi, respectively, along a high-symmetry momentum trajectory, measured at $T$ = 40 K. Data measured with $E_{\rm i}=57$ meV and 90 meV are combined after proper intensity normalization. (d) The Brillouin zone (BZ), with the momentum trajectory marked in red. (e) and (f) $\cal S(\mathbf{Q},\omega)$ calculated from the fitted-force-constant model along the same trajectories as in (b) and (c). Solid lines in (b), (c), (e), and (f) indicate the calculated phonon dispersions.}
\end{figure*}

It is experimentally possible to determine Chern numbers in phonon and magnon bands, if one can measure the topological structure of wave functions, \textit{i.e.}, eigenvectors of motion, in momentum space. Scattering methods such as inelastic neutron scattering (INS) and X-ray scattering are suitable for this purpose, because their dynamical structure factor $\cal S(\mathbf{Q},\omega)$ is related to the excitations' eigenvectors \cite{Lovesey1984,squires_2012}. As the eigenvectors vary strongly in the vicinity of Weyl and Dirac points, the observed intensities are expected to undergo strong and distinct modulations, which can reflect the topological structure. Such modulations have been recently reported in several topological magnon systems around Dirac points \cite{Shivam,Elliot2021} and nodal lines \cite{Scheie2022,Scheie202202}.

Here, we report an INS study of $M$Si ($M=$ Co, Mn) single crystals, which host multiple types of topological phonon band crossing nodes \cite{Zhangtiantian2018, Zhangtiantian2020, Miaohu2018}. By comparing the observed INS intensities with our fitted model based on density functional perturbation theory (DFPT) calculations,  we verify the theoretically predicted coexistence of two-fold quadruple Weyl points, three-fold spin-1 Weyl points, and four-fold charge-2 Dirac points in $M$Si. We further explore the spectroscopic signatures of topological structures near the topological band crossing points and show, theoretically and in some cases with comparison to the experimental data, that the number of intensity extrema on a momentum sphere enclosing the band-crossing node equals the Chern number of the node. Our result demonstrates the capability of INS for direct Chern-number determination.

This paper is organized as follows: In Section II, we describe the INS experiment and the fitted-force-constant model based on DFPT. In Sec. III, we show phonon dispersions, both in a global view and close to topological band crossing nodes. In Sec. IV, we discuss INS spectroscopic features near the band-crossing nodes and investigate their relation with the Chern numbers. In Sec. V, we make a brief discussion and a summary.

\section{Experiment and calculation methods}
\subsection{INS experiment}
High-quality single crystals of MnSi and CoSi were grown by a traveling-solvent floating zone method. The INS experiments were performed on the 4SEASONS spectrometer at MLF, J-PARC, Japan and the SEQUOIA spectrometer at SNS, ORNL, USA \cite{season, Granroth_2010}. A total of 33 (28) grams of MnSi (CoSi) twin-free single crystals with a mosaic spread of $\lesssim1.3^\circ$ full-width at half-maximum (FWHM) were used for the experiments (Fig.~\ref{figs1} \cite{sm,jin2022}). The INS data shown in this paper were collected with incident neutron energies $E_{\rm i}$ = 57 and 90 meV at a fixed temperature of $T$ = 40 K, and analyzed with the \textsc{utsusemi} and \textsc{horace} software \cite{Utsusemi,EWINGS2016132}. As MnSi and CoSi share the $B$20-type structure belonging to the non-centrosymmetric space group $P2_13$ [Fig.\ref {fig1}(a)], data from equivalent momenta have been symmetrized and averaged accordingly in order to improve counting statistics. Intensities are presented in absolute scattering cross sections by using the incoherent elastic scattering of the sample for normalization \cite{Xuguangyong}. To best visualize the phonon cross sections, we present coherent scattering signals from the sample only, whereas intensities arising from incoherent scattering of the sample and from the aluminum sample holders have been subtracted as background (Fig.~\ref{figs2} \cite{sm}).

\begin{figure*}
\centering
\includegraphics[width=\textwidth]{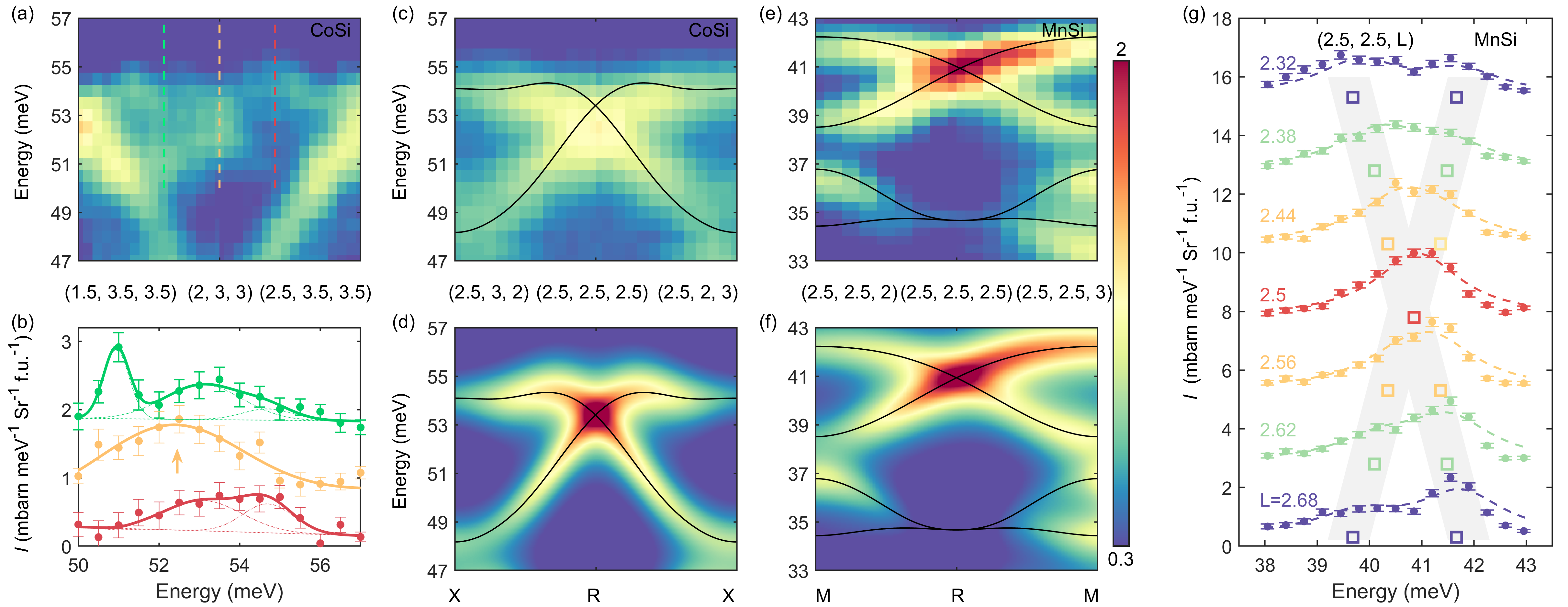}
\caption{\label{fig2} (a) INS spectra near a spin-1 Weyl point in CoSi, plotted along a R-$\Gamma$-R trajectory. Colored dashed lines correspond to energy cuts in (b), which are fitted with a sum of Gaussian profiles on a linear background, by assuming a total of one and three peak(s) at and away from the $\Gamma$-point, respectively, and under the constraint that equivalent $\mathbf{q}$ positions must have the same energies and peak widths. (c) and (d) Phonon intensities from INS experiment and the fitted-force-constant model, respectively, plotted along a X-R-X trajectory near a charge-2 Dirac point in CoSi. (e) and (f) Similar to (c) and (d), but for MnSi and along an M-R-M trajectory. (g) Energy cuts at a series of $\mathbf{Q}$ points using the same data as in (e) and (f). The INS and fitted model calculated data are displayed by circles and dashed lines, respectively. Open squares denote peak positions estimated from two-peak fits to the data (the fits are not shown), which form an approximate linear band crossing.  Data in (b) and (g) are offset for clarity.}
\end{figure*}

\subsection{Fitted-force-constant model}

The phonon force constant matrices of CoSi and MnSi were calculated with the Vienna \textit{ab intio} simulation package (\textsc{vasp}) \cite{Gonze1997,Kresse1993,Kresse1994,Kresse1996,KresseG1996} using the DFPT method. The calculations were done with the Perdew-Burke-Ernzerhof (PBE) type exchange-correlation functional, under the generalized gradient approximation (GGA) \cite{Perdew1996}. The kinetic energy cutoff was set to 400 eV. Integrations over the Brillouin zone were performed with Monkhorst-Pack $\mathbf{Q}$-point grids (equivalent to $12\times 12\times 12$ grid for CoSi and $15\times 15\times 15$ grid for MnSi). Lattice constants and atomic positions were relaxed until residual forces drop below 0.001 eV/\AA. The relaxed lattice constants were 4.35~\AA{} for CoSi and 4.42~\AA{} for MnSi, which are slightly smaller than our experimental values, $4.43\pm 0.02$~\AA{} for CoSi and $4.56\pm 0.01$~\AA{} for MnSi. After obtaining the band dispersion $\omega(\mathbf{Q})$, the coherent dynamical structure factors ${\cal S}_{\text{coh}}^{(i)}(\mathbf{Q},\omega)$ for all vibration modes $(i)$ were written as 
 \cite{Lovesey1984,squires_2012,shirane_shapiro_tranquada_2002}
\begin{align}\label{eq01}
    \begin{split}
       {\cal S}_{\text{coh}}^{(i)}(\mathbf{Q},\omega)&= \frac{(2\pi)^3}{v_0}\sum_{\mathbf{G}, \mathbf{q}}  \frac{|F^{(i)}_{\text{coh}}(\mathbf{Q})|^{2}}{2\omega^{(i)}(\mathbf{q})} \delta(\mathbf{Q}-\mathbf{q}-\mathbf{G})\delta(\omega-\omega^{(i)}(\mathbf{q})),\\
 F^{(i)}_{\rm coh}({\bf Q})&= \sum_d\frac{b_{d,\text{coh}}}{\sqrt{m_d}}\mathbf{Q}\cdot\bm{\xi}^{(i)}_d(\mathbf{q}) e^{i\mathbf{Q}\cdot \mathbf{r}_d},
    \end{split}
\end{align}
where $\mathbf{Q}=\mathbf{q}+\mathbf{G}$ is the total momentum transfer with $\mathbf{G}$ being a reciprocal lattice vector, and $m_d$, $\mathbf{r}_d$, $b_{d,\text{coh}}$, $\bm{\xi}^s_d(\mathbf{q})$ denote the mass, position, coherent scattering length and eigen vector of the $d$\textsuperscript{th} atom in the primitive cell.

Notably, the original DFPT results have deviations from the INS data, including a global rescaling in energies, slight distortions in the dispersions and scattering intensity (Fig.~\ref{figs5} \cite{sm}). Such deviations may stem from inaccuracy in the calculated crystal structure and force constants. To improve the accuracy of our model, we have performed parametric fits on the leading force constants while discarding the weaker interactions. The energy values of all phonon branches at high-symmetry points ($\Gamma$, X, M, and R) are extracted from the experimental spectra (Table ~\ref{tab1} \cite{sm}) and used in the fitting. By adopting about ten pairwise interactions, we are able to reproduce features in the experimental dispersions and intensities with satisfactory accuracy over many BZs (Fig.~\ref{figs5} \cite{sm}). All model calculation results in the main text are obtained with the optimized parameters listed in Table ~\ref{tab2} \cite{sm}.

\section{Dispersion and topological band crossing}
To begin, we present in Fig.~\ref{fig1}(b) and (c) representative INS spectra of MnSi and CoSi, respectively, along high-symmetry lines in the irreducible Brillouin zone (BZ) [Fig.~\ref{fig1}(d)]. A wealth of phonon scattering signals are observed. Strong contaminations below 10 meV are due to multiple scattering, and residual aluminum phonon scattering are observed below 25 meV (especially in the CoSi data). Overall, the phonon INS signals compare very well with our model calculations in Fig.~\ref{fig1}(e) and (f). The model for MnSi is particularly satisfactory. This suggests that our fitted-force-constant model provides a good approximate representation of the phonons.

Having established our fitted model, we now zoom into the topological band crossings. We start from the $\Gamma$-point at the BZ center. According to group theoretical analysis, phonons at the $\Gamma$-point are irreducible representations of the tetrahedral $T(23)$ point group: $\Gamma = 2A + 2E + 6T$, where $A$, $E$, and $T$ represent singly, doubly, and triply degenerated modes, respectively. Importantly, all triply degenerate phonons at the $\Gamma$-point are spin-1 Weyl points protected by the point-group symmetry \cite{Zhangtiantian2018}, and all doubly degenerate phonons are quadruple Weyl points protected by the extra time-reversal symmetry $\mathcal{T} $ \cite{Zhangtiantian2020, Liu2020}. We have additionally verified the irreducible representations of the BZ-center phonons using polarized Raman spectroscopy (Fig.~\ref{figs3} \cite{sm}).

Based on the above information, we zoom into the spin-1 Weyl point at the highest energy (52.5 meV) in CoSi. The INS intensities along a R-$\Gamma$-R momentum trajectory are displayed in Fig.~\ref{fig2}(a). This band crossing has relatively large dispersion velocities and is far away from other bands, yet still, the fact that a total of three branches are involved in the crossing makes them challenging to resolve experimentally. By making energy line cuts and fitting the intensity profiles systematically [Fig.~\ref{fig2}(b)], we find that the INS data do support a crossing of three bands. For comparison, fitting the spectra obtained symmetrically away from $\Gamma$ with only two or fewer peaks does not yield a consistent description (Fig.~\ref{figs4}(a) \cite{sm}). We thus conclude that the spin-1 Weyl point in CoSi is experimentally confirmed. In MnSi, however, all three-fold degenerate modes at the $\Gamma$-point are too close in energy to other phonon branches [Fig.\ref{fig1}(b) and (e)], precluding a similar confirmation to be made.


We next turn to the R-point at the BZ corner, where all band crossings are four-fold degenerate ``charge-2 Dirac points'' ensured by the crystallographic and $\mathcal{T}$ symmetries \cite{Zhangtiantian2018}. A charge-2 Dirac point is the direct sum of two identical spin-1/2 Weyl points. Because the bands remain two-fold degenerate along the $\overline{\rm RX}$ and $\overline{\rm RM}$ directions, we expect to observe only two linearly dispersing branches along these directions, which can be regarded as a key signature of the Dirac points. INS spectra consistent with such understanding are displayed near 41 meV for MnSi and 52 meV for CoSi in Fig.~\ref{fig2}(c) and (e), respectively. The band crossings are also reproduced in our model calculations [Fig.~\ref{fig2}(d) and (f)]. Energy cuts at a series of successive $\mathbf{Q}$ positions [Fig.~\ref{fig2}(e)] further confirm the approximate linear band crossing [Fig.~\ref{fig2}(g)], and a similar case for CoSi is displayed in Fig.~\ref{figs4}(b) \cite{sm}. For MnSi, the fitted-force-constant model is furthermore able to account for the scattering cross sections quite accurately, as seen from the colored dashed lines in Fig.~\ref{fig2}(g), which actually represent model calculated intensities rather than peak fitting. This quantitative agreement suggests that it is possible to use the intensity information, from the experiment and/or the calculation, to elucidate the Chern number of a topological band crossing, which is our next subject.

\section{Detection of Chern numbers by neutron scattering}
\subsection{General theoretical scheme for two-fold Weyl points}
In this section, we will derive the explicit relation among phonon eigenvectors, Chern numbers, and the INS dynamical structure factor, using low-energy effective models near topological band crossing points. We will first present theoretical considerations for two-fold Weyl points. Then, we will use the general formulism in our specific analyses of four-fold charge-2 Dirac points ($2\times$ two-fold Weyl points) and quadruple Weyl points.

To begin with, for a two-fold Weyl point, the effective Hamiltonian can be written as a $2\times 2$ Hermitian matrix
\begin{align}
H_{2\times 2}(\mathbf{q})=\sum_{i=x,y,z}{f_i}(\mathbf{q})\cdot {\sigma_i} +f_0(\mathbf{q})\sigma_0,
\end{align}
where $\mathbf{q}$ is momentum measured from the Weyl point and ${\sigma_i}$ are the Pauli matrices. In this notation, the eigenvector of one of the bands (\textit{e.g.}, the upper band) is represented by a spinor $\bm{\xi}^{\text{up}}(\mathbf{q})=(\psi_1(\mathbf{q}),\psi_2(\mathbf{q}))^{\text{T}}$. We can further define a pseudospin quantity $\mathbf{S}({\mathbf{q}})$
\begin{align}
S_{i}(\mathbf{q}) \equiv \frac{f_{i}(\mathbf{q})}{|\mathbf{f}(\mathbf{q})|}=\left\langle\xi^{\text {up}}\left|\sigma_{i}\right|\xi^{\text {up}}\right\rangle,
\end{align}
whose direction is represented by a point on the Bloch sphere $\mathbb{S}_{\mathrm B}$. If we consider a surface $\mathbb{S}_{\mathbf{q}}$ enclosing the Weyl point in momentum space, on which a gap always exists between the upper and lower bands, a wrapping number can be used to characterize the mapping from $\mathbb{S}_{\mathbf{q}}$ to $\mathbb{S}_{\mathrm B}$. This number is the Chern number $\pm C$ of the Weyl node \cite{Zhangtiantian2020}, which does not depend on the shape of $\mathbb{S}_{\mathbf{q}}$. Specifically, $\mathbf{S}({\mathbf{q}})$ will take every possible directions on $\mathbb{S}_{\mathrm B}$ at least $|C|$ times as $\mathbf{q}$ moves around $\mathbb{S}_{\mathbf{q}}$. We will see examples of this in Figs.~\ref{fig3}(b) and \ref{fig4}(a).

Next, we show that the pseudospin texture $\mathbf{S}({\mathbf{q}})$ can leave distinct signatures in the INS dynamical structure factors. As we can see from Eq.~(\ref{eq01}), the dynamic structure factor ${\cal S}_{\text{coh}}^{(i)}(\mathbf{Q},\omega)$ is sensitive to the inner product of $\bm{\xi}^{(i)}_d(\mathbf{q})$ and $\mathbf{Q}$. Taking acoustic phonons for example, the INS intensity vanishes if the polarization vector $\bm{\xi}$ (same for all atoms for acoustic phonons) lie perpendicular to $\mathbf{Q}$, and reaches maximum when the two vectors are parallel. For optical branches, similar conclusions do exist by generalizing the real-space polarization vectors to abstract phonon eigenvectors in the Hilbert space. The main difference is that the eigenvector now has $3 N$ distinct components (for describing the collective vibration of $N$ atoms in the primitive cell) which are generically complex numbers. In the close vicinity to a Weyl node, assuming that $|\mathbf{Q}|\gg |\mathbf{q}|$ and $|\mathbf{q}|\ll 2\pi/a$, so that $\mathbf{Q}\approx \mathbf{G}$, the only fast varying term in the formula is the phonon eigenvector $\bm{\xi}^{\text{up}}(\mathbf{q})$. Using the pseudospin quantities, it is straightforward to show that the INS intensity of the upper band is approximately \cite{sm}
\begin{align}\label{eqnfinal}
{\cal S}_{\text{coh}}^{\rm up}(\mathbf{Q},\omega)&\propto |\mathbf{V}| \left(1+\frac{\mathbf{S}(\mathbf{q}) \cdot \mathbf{V}(\mathbf{G})}{ \left|\mathbf{S}(\mathbf{q})\right| \left|\mathbf{V}(\mathbf{G})\right|} \right),
\end{align}
where $\mathbf{V}(\mathbf{G})$ is a constant vector that does not sensitively depend on $\mathbf{q}$ but varies between BZs. Both the pseudospin $\mathbf{S}$ and the vector $\mathbf{V}$ are vectors in the effective two-band Hilbert space, which ultimately encode the phonon eigenvectors and their sampling by INS in the BZ of $\mathbf{G}$. It then becomes clear that the INS intensity is related to a projection of $\mathbf{S}(\mathbf{q})$ along $\mathbf{V}$. In particular, ${\cal S}_{\text{coh}}^{\rm up}(\mathbf{Q},\omega)$ would reach maximum when $\mathbf{S}(\mathbf{q})$ and $\mathbf{V}$ are parallel, and vanish when they are anti-parallel.

Combining Eq.~(\ref{eqnfinal}) with our former arguments about the wrapping (Chern) number, we come to the following explicit statement: On a momentum surface $\mathbb{S}_{\mathbf{q}}$ that encloses the Weyl node, there are at least $|C|$ momenta where pseudospin $\mathbf{S}$ is parallel to $\mathbf{V}$, where the INS intensity of the upper band ${\cal S}_{\text{coh}}^{\rm up}(\mathbf{Q},\omega)$ reaches maximum. Similarly, there are also at least $|C|$ momenta where the intensity reaches zero. In most practical cases where the lowest-order $k\cdot p$ theory holds, the total numbers of maxima and zeros are simply $|C|$. The approximation of $\mathbf{Q}=\mathbf{q}+\mathbf{G}\approx \mathbf{G}$ can also be released since $\mathbf{Q}$ has trivial topology on the surface $\mathbb{S}_{\mathbf{q}}$ as long as the origin of $\mathbf{Q}$ is not enclosed. In summary, a Weyl node serves as a singular point of the pseudospin in momentum space, and the contrasting INS intensity distribution around it reveals its Chern number.

\subsection{The charge-2 Dirac point}

\begin{figure}
\includegraphics[width=8.6cm]{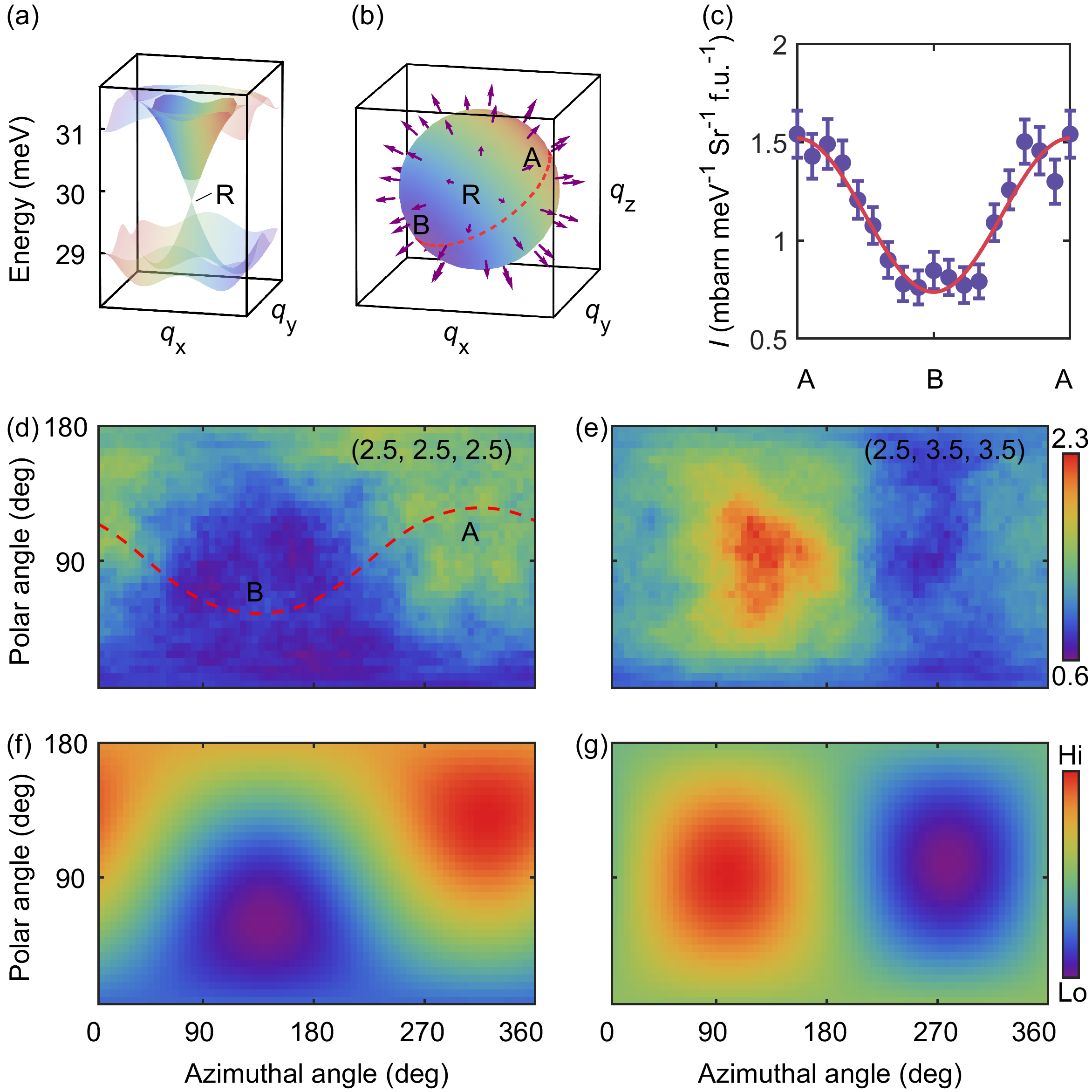}
\caption{\label{fig3} (a) Phonon dispersions near the charge-2 Dirac point at 29.6 meV in MnSi. Solid color indicates energy integration range (30--32 meV). (b) Pseudospin texture on a $\mathbf{q}$-sphere enclosing the R-point. Color indicates energy-integrated INS intensities near $\mathbf{Q}=(2.5,\,2.5,\,2.5)$. (c) INS intensity along the red circle $\overset{\frown}{ABA}$ in (b), fitted with a cosine function (red line, see text). (d) and (e) INS intensities averaged over solid $\mathbf{q}$-spheres enclosing $(2.5,\,2.5,\,2.5)$ and $(2.5,\,3.5,\,3.5)$ respectively, with a radius of 0.2 r.l.u. and a cone-smoothing width of $\pm20^{\circ}$. (e) and (f) Fitted-force-constant model calculations according to Eq.~(\ref{eqnfinal}), for the same $\mathbf{q}$-spheres as in (d) and (e), respectively.}
\end{figure}

Now that we have linked the pseudospin texture with the INS intensity distribution, we next use it to analyze the charge-2 Dirac point in the phonon bands of $M$Si at the R-point of the BZ. Although being four-fold degenerate, the effective Hamiltonian near the band crossing
$H_4(\mathbf{q})\propto\left(\begin{smallmatrix}
\mathbf{q}\cdot \bm{\sigma} & 0\\
0 & \mathbf{q}\cdot \bm{\sigma}
\end{smallmatrix}\right)$
is the direct sum of that of two identical spin-1/2 Weyl points, each with Chern number $C=\pm 1$\cite{Zhangtiantian2018}. The general statement between pseudospin and INS intensity can be easily generalized in this case with only minor modifications \cite{sm}. Namely, as long as the INS intensities of the two upper bands are considered as a whole, there will be exactly one maximum and one minimum on the enclosing $\mathbb{S}_{\mathbf{q}}$, \textit{i.e.}, resembling that of a regular spin-1/2 Weyl point. The only difference is an extra constant term in the intensity, so that the minimum is finite rather than zero \cite{sm}.

For a concrete example, we inspect the INS intensities near a charge-2 Dirac point (at the BZ R-point) at about 29.6 meV in MnSi. The phonon dispersion nearby is schematically shown in Fig.~\ref{fig3}(a). On a small $\mathbf{q}$-sphere around the R-point, the pseudospin texture is visualized in Fig.~\ref{fig3}(b), where purple arrows indicate the pseudospin directions $\mathbf{S}(\mathbf{q})$. The outward hedgehog configuration of the arrows indicates that $\mathbf{S}(\mathbf{q})$ takes every direction once on the Bloch sphere, \textit{i.e.}, the Chern number is $|C|=1$.

From an experimental perspective, a $\mathbf{q}$-sphere as small as possible should be used to extract the INS intensity, in order to avoid an overlap with neighboring bands. But the sphere cannot be too small or thin, as otherwise the counting statistics would be too low. In our case, we find that binning INS data over a finite solid sphere produces satisfactory results. In Fig.~\ref{fig3}(d), intensities integrated over [30, 32] meV around R = (2.5, 2.5, 2.5) are displayed as a function of the polar and azimuthal angles. The radius of the solid sphere is set to be 0.2 reciprocal lattice units (r.l.u.), and the displayed intensity at each angle represents the average over a cone volume within a $20^{\circ}$ half-apex angle.

The above INS result agrees nicely with our fitted-force-constant model calculations [Fig.~\ref{fig3}(f)]. To rationalize their characteristics with our effective model in Eq.~(\ref{eqnfinal}), we note that due to the three-fold rotational symmetry, the vector ${\mathbf{V}}$ for the Dirac point at R = (2.5, 2.5, 2.5) is along the $[111]$ direction. Consequently, the intensity will reach its maximum and minimum along the $[111]$ and $[\bar{1}\bar{1}\bar{1}]$ directions (or vice versa). Moreover, on a great circle $\overset{\frown}{ABA}$ [red dashed lines in Fig.~\ref{fig3}(b) and (d)] that passes through the $[111]$ and $[\bar{1}\bar{1}\bar{1}]$ directions, the intensity is expected to have a cosine behavior, which we also confirm experimentally [Fig.~\ref{fig3}(c)].

Besides for the R-point (2.5, 2.5, 2.5) where the symmetry is high, the same analysis can be performed at other R-points, such as (2.5, 3.5, 3.5). The results [Fig.~\ref{fig3}(e) and (g)] show that the extrema are no longer located along the diagonal direction, due to the different orientation of $\mathbf{V}$. Nevertheless, the fact that the intensity exhibits one minimum and one maximum indicates that the underlying Chern number is $|C|=1$.

\subsection{The quadruple Weyl point}

\begin{figure}
\includegraphics[width=8.6cm]{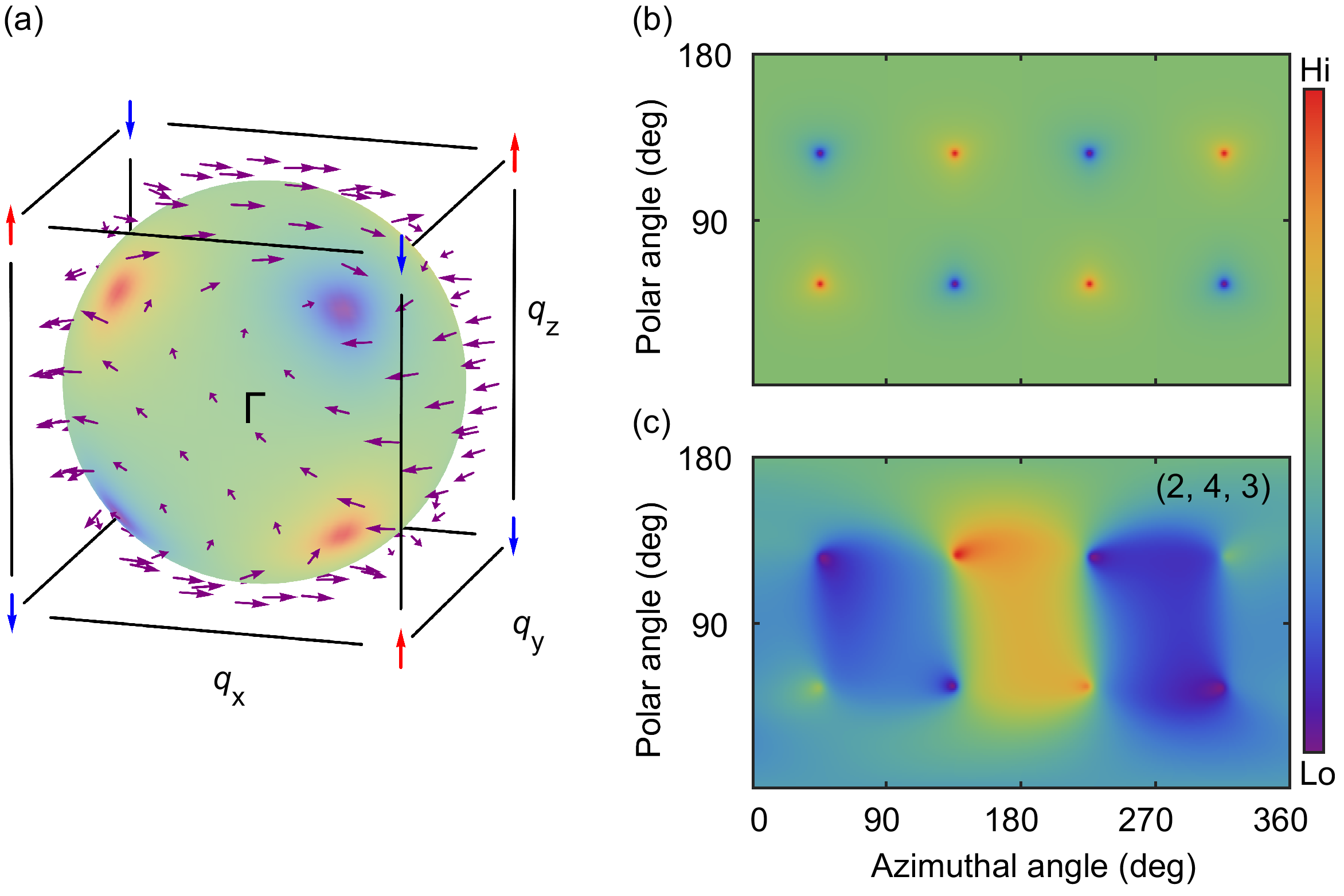}
\caption{\label{fig4} (a) The pseudospin texture on a $\mathbf{q}$-sphere enclosing $\Gamma$, with the colors indicating the magnitude of $S_z$ component. The orientations of $\mathbf{S}$ at 8 diagonal directions are marked especially by blue and red arrows, showing tetrahedral symmetry. (b) $S_z$ component of the pseudospin as a function of polar and azimuthal angles on the sphere. (c) Simulated INS intensity near $\mathbf{Q}=(2, 4, 3)$, showing similar patterns to (b) since $\mathbf{V}$ is almost in the $z$ direction. All simulations are done with $|\mathbf{q}|=0.05$ r.l.u. and for the upper band near the 40.1 meV band-crossing point in MnSi.}
\end{figure}

We further explore the quadruple Weyl points at the $\Gamma$-point, which are protected by time-reversal symmetry $\mathcal{T}=K \sigma_x$ and have an unusual Chern number of $\pm 4$. The effective Hamiltonian can be written as \cite{Zhangtiantian2020}
\begin{equation}
H_{2\times 2}(\mathbf{q})=-\begin{pmatrix}
A q_{x} q_{y} q_{z} & B\left(q_{x}^{2}+\omega q_{y}^{2}+\omega^{2} q_{z}^{2}\right) \\
B\left(q_{x}^{2}+\omega^{2} q_{y}^{2}+\omega q_{z}^{2}\right) & -A q_{x} q_{y} q_{z}
\end{pmatrix},
\end{equation}
where $\omega=\exp (-2\pi i/3)$, and $A$ and $B$ are real constants. We have omitted the kinetic energy term $f_0(\mathbf{q})$ that has nothing to do with the band topology. The pseudospin texture $\mathbf{S}(\mathbf{q})$ around the quadruple Weyl point is shown in Fig.~\ref{fig4}(a). The $S_z$ component of the pseudospin is illustrated by the colors on the spherical $\mathbf{q}$-surface in Fig.~\ref{fig4}(a) and plotted on a flat map in Fig.~\ref{fig4}(b). The direction of $\mathbf{S}(\mathbf{q})$ is almost always in the $q_xq_y$ plane except near $[111]$ and equivalent diagonal directions. As indicated by the blue and red arrows in Fig.~\ref{fig4}(a), the eight diagonal directions can be sub-divided into two tetrahedron-vertex sets with opposite $S_z$, and the total wrapping number (Chern number) is $|C|= 4$.

An experimentally unfavorable aspect of the quadruple Weyl point is in the weak dispersion -- the energy splitting between the two bands only increases as a quadratic (in most directions) or even a cubic (along $[111]$ and its equivalents) function of $\mathbf{q}$. This makes it difficult to separately measure ${\cal S}(\mathbf{Q},\omega)$ of one of the bands. Moreover, all the quadruple Weyl points in MnSi and CoSi turn out to be close to other phonon branches. With the energy resolution of our INS experiment, we are unable to extract the pseudospin texture of the quadruple Weyl points from the INS data. Nevertheless, our fitted numerical model have no such resolution limits. In Fig.~\ref{fig4}(c), we calculate ${\cal S}(\mathbf{Q},\omega)$ of the upper band of the quadruple Weyl point at 40.1 meV in MnSi, on a 0.05 r.l.u. $|\mathbf{q}|$-sphere surrounding $\Gamma = (2, 4, 3)$. The BZ center $(2, 4, 3)$ is chosen because the $\mathbf{V}$ vector lies very close to the $z$ direction, such that the INS intensities on the $\mathbf{q}$-sphere exhibit a similar distribution as the $S_z$ component in Fig.~\ref{fig4}(b), with four maxima and four minima approximately along the $\langle{111}\rangle$ diagonal directions. This ``virtual'' measurement suggests that INS has in principle the capability to reveal the pseudospin's wrapping behavior (and hence the Chern number) associated with the novel quadruple Weyl points. It can be realized in experiments with higher resolution and/or in materials where the quadruple Weyl points are far away from other phonons.

\section{Discussion and conclusion}

Our study demonstrates the capability of INS to measure Chern numbers of topological phonon band crossing. While scattering intensity can vary with momentum even on non-topological bands, the unique spectroscopic characteristics of Weyl and Dirac points lie in the fact that they are singular points for eigenvectors, hence the intensity distribution around them exhibits abrupt variations: Even on an infinitely small enclosing momentum surface, the intensity modulations around the topological nodes are still present, whereas in the case of topologically trivial band crossings, the intensity distribution would approach a constant as the enclosing surface shrinks into a point. Consequently, in order to determine Chern numbers from INS experiments, it is better to study the close vicinity of the band crossing points, \textit{i.e.}, using a small $\mathbf{q}$-surface to both avoid other bands and ensure that the intensity modulations arise solely from the topology. We also note that the intensity modulations we discuss here are universal for two-fold phonon Weyl points of any Chern number, as well as for some of the four-fold Dirac points. According to group theory \cite{Liu2020}, phonon Weyl points exist in crystals of certain space groups, and they can be found at specific high-symmetry points in the BZ. As long as they are far away from other bands and have a relatively large group velocity, intensity modulations should be observable by INS.

The detection of wave functions (or, vibrational eigenvectors) is not restricted to INS experiments on phonons, and it can be a common capability of many spectroscopic methods. In polarization-dependent angle-resolved photoemission spectroscopy (ARPES), changes in the signal intensity has been suggested to reflect the wave functions of Dirac electrons \cite{Hwang2011}. Related measurements have also been proposed for resonant inelastic x-ray scattering (RIXS) \cite{Kourtis2016}. While the wave-function texture of the quasiparticles is at the origin of all the spectroscopic observables, the specific interactions between the experimental probes and the quasiparticles may add further complexity to the interpretation of experiments. For instance, INS measurements of phonons involve neutron collisions with nuclei, giving rise to the $\mathbf{Q}\cdot \bm{\xi}$ term in the scattering cross section; and because $\mathbf{Q}$ uniquely determines $\mathbf{V}$ in Eq.~(\ref{eqnfinal}), the associated projection of $\bm{\xi}$ always allows for the determination of the Chern number. On the contrary, for magnon bands, dipole-dipole interactions between neutrons and magnetic moments lead to a $\mathbf{Q}\times (\mathbf{Q}\times \mathbf{S})$ term in the cross section, and the component of $\mathbf{S}$ parallel to $\mathbf{Q}$ is missing from the detection. As a result, the number of extrema in the intensity modulation on an enclosing momentum surface may change between different choices of the measurement BZ, rendering it necessary to have extra knowledge about the magnetic system in order to correctly infer the topological invariant. In the cases of ARPES and RIXS, complexity may arise because the different polarization channels have to be considered together. To this end, our INS measurement of phonons may be regarded as a demonstration of principles that motivates further studies.


In conclusion, we have performed a comprehensive study on the topological phonon band crossings in MnSi and CoSi. Both the band dispersions and the coherent dynamical structure factors are experimentally resolved with high precision, yielding results that compare well with model calculations. The existence of spin-1 Weyl points and charge-2 Dirac points in the phonon bands are verified by the dispersions. Combining experiments and model calculations, we further demonstrate the capability of INS for unambiguously determining Chern numbers of band crossing nodes. Our general theoretical scheme based on effective Hamiltonians suggests that related methods can be found in the study of other topological quasiparticles, as well as in the use of other spectroscopic methods.

\begin{acknowledgments}
The work at Peking University was supported by the National Key R\&D Program of China (No. 2018YFA0305602) and the National Natural Science Foundation of China (Nos. 12061131004 and 11888101).  The work at Brookhaven National Laboratory was supported by Office of Basic Energy Sciences (BES), Division of Materials Sciences and Engineering, U.S. Department of Energy (DOE), under contract DE-SC0012704. Part of this research was performed at the MLF, J-PARC, Japan, under a user program (proposal Nos. 2018A0193, 2018B0201, 2019A0085). A portion of this research used resources at Spallation Neutron Source, a DOE Office of Science User Facility operated by the Oak Ridge National Laboratory.
\end{acknowledgments}
\normalem
\bibliography{MSi_phonon_refs.bib}
\clearpage
\widetext
\begin{center}
\textbf{\large Supplemental Material for\\ ``Chern numbers of topological phonon band crossing determined with inelastic neutron scattering''}
\end{center}
\renewcommand{\thetable}{S\arabic{table}}
\renewcommand{\thefigure}{S\arabic{figure}}
\renewcommand{\theequation}{S\arabic{equation}}
\renewcommand{\tablename}{TABLE}
\section{Details in experiment}
  \begin{figure}[!ht]
    \includegraphics[width=0.7\textwidth]{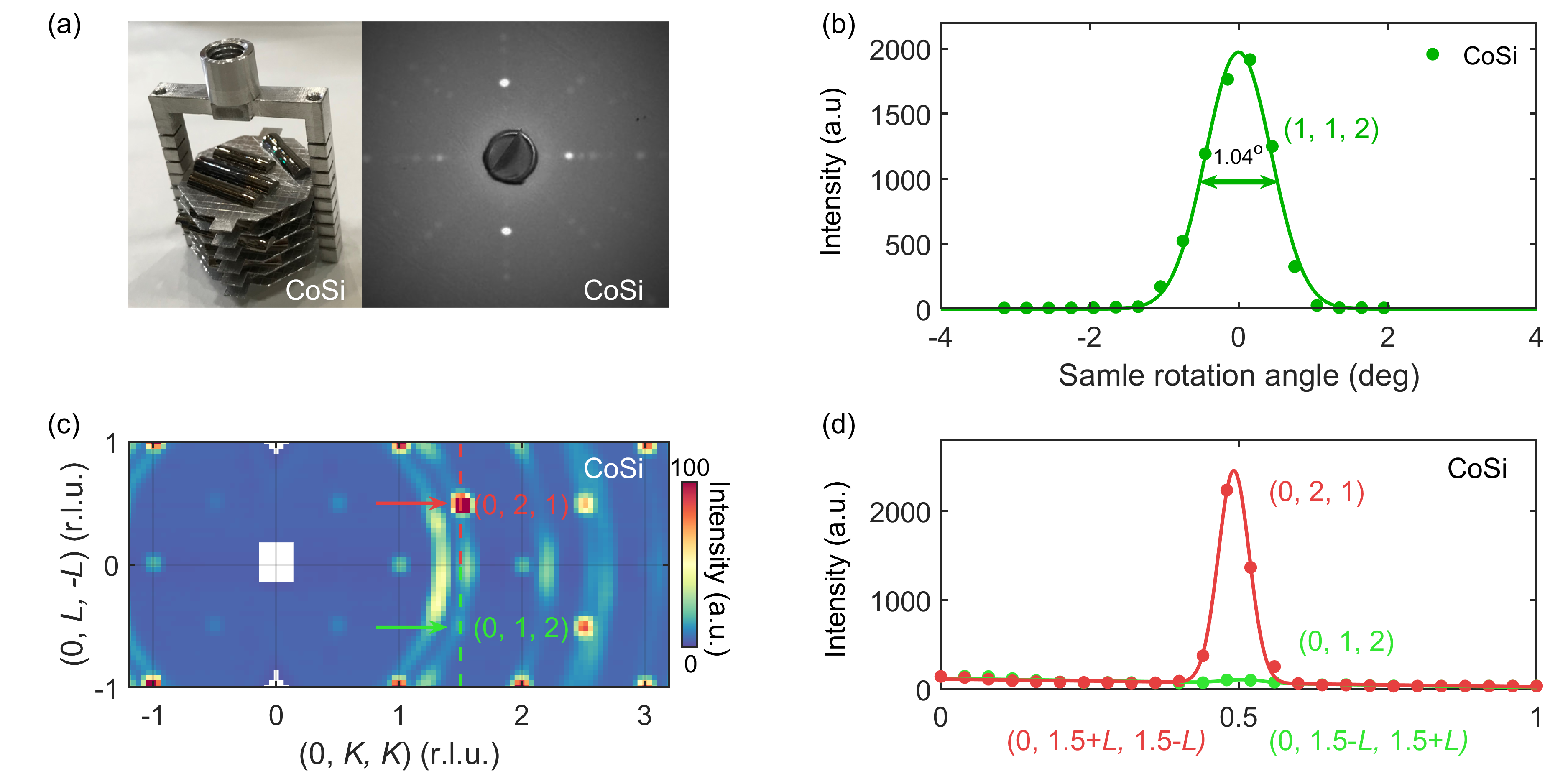}
    \caption{\label{figs1}(a) Left side: photograph of CoSi single crystals co-aligned on an aluminum sample holder. Right side: representative X-ray Laue pattern taken on a natural surface of one CoSi single crystal along $\langle 100\rangle$ direction. (b) Neutron diffraction intensities of the (1, 1, 2) Bragg reflection recorded when rotating the entire sample array. The solid line is a Gaussian fit to the data, with full-width-half-maximum (FWHM) of 1.04${}^{\circ}$. (c) and (d) $\mathbf{Q}$-scan profiles of CoSi indicating that the entire sample is nearly single-domained, since the (0, 2, 1) reflection is over 25 times more intense than that of (0, 1, 2).}
  \end{figure}
  \begin{figure}[!ht]
    \includegraphics[width=0.7\textwidth]{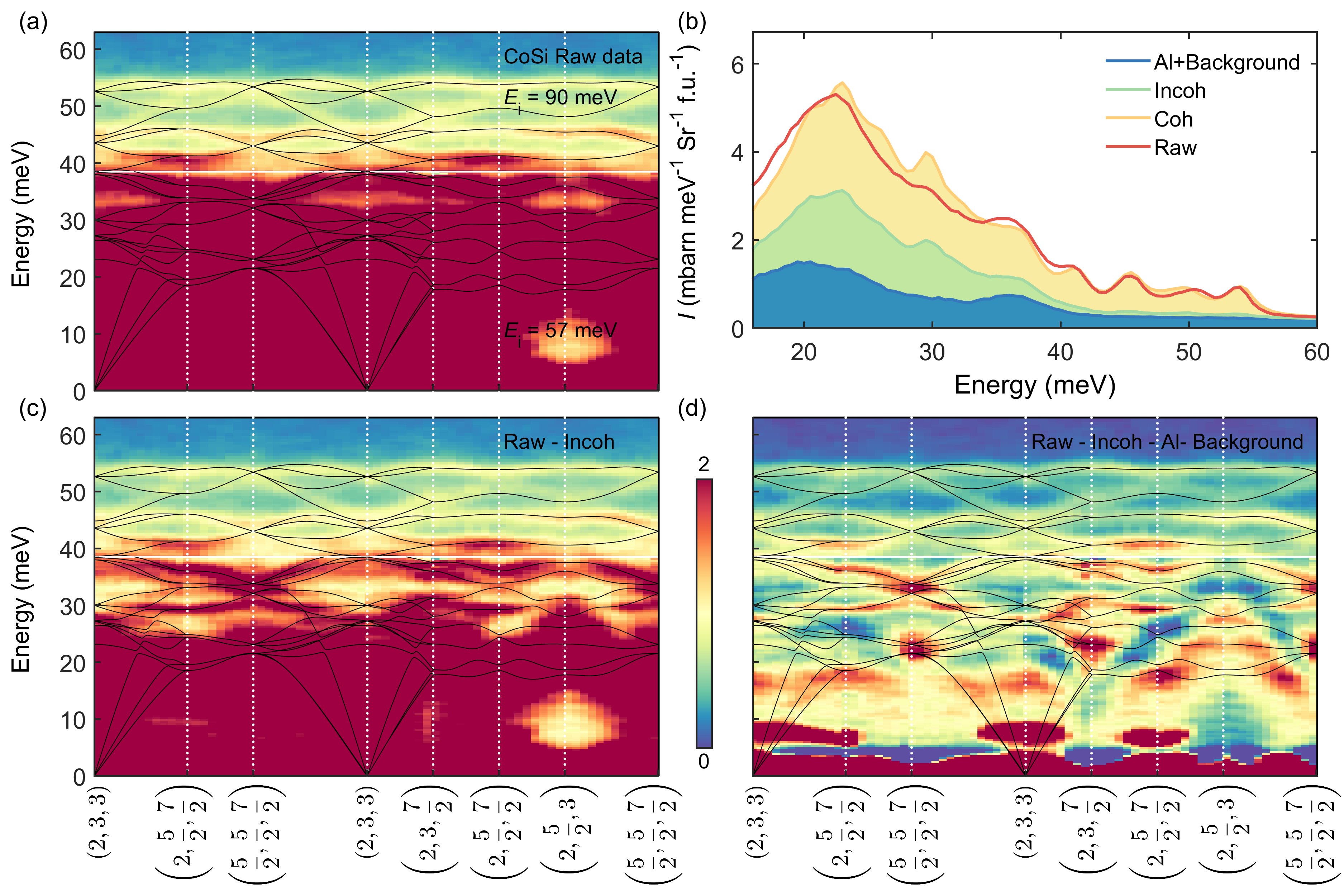}
    \caption{\label{figs2}(a) Raw INS spectra of CoSi along the chosen trajectory of high symmetry directions. (b) $\mathbf{Q}$-averaged INS intensity over the entire $(2,3,3)$ BZ. The contributions of calculated coherent, incoherent and fitted aluminum (with background) scattering intensities are shown respectively. (c) Modified INS spectra after subtracting incoherent signals. (d) Best-estimated coherent INS spectra shown in the main text, obtained by subtracting both incoherent and aluminum signals.}
    \vspace{0.5cm}
  \end{figure}

  \begin{figure}[!ht]
    \includegraphics[width=0.7\textwidth]{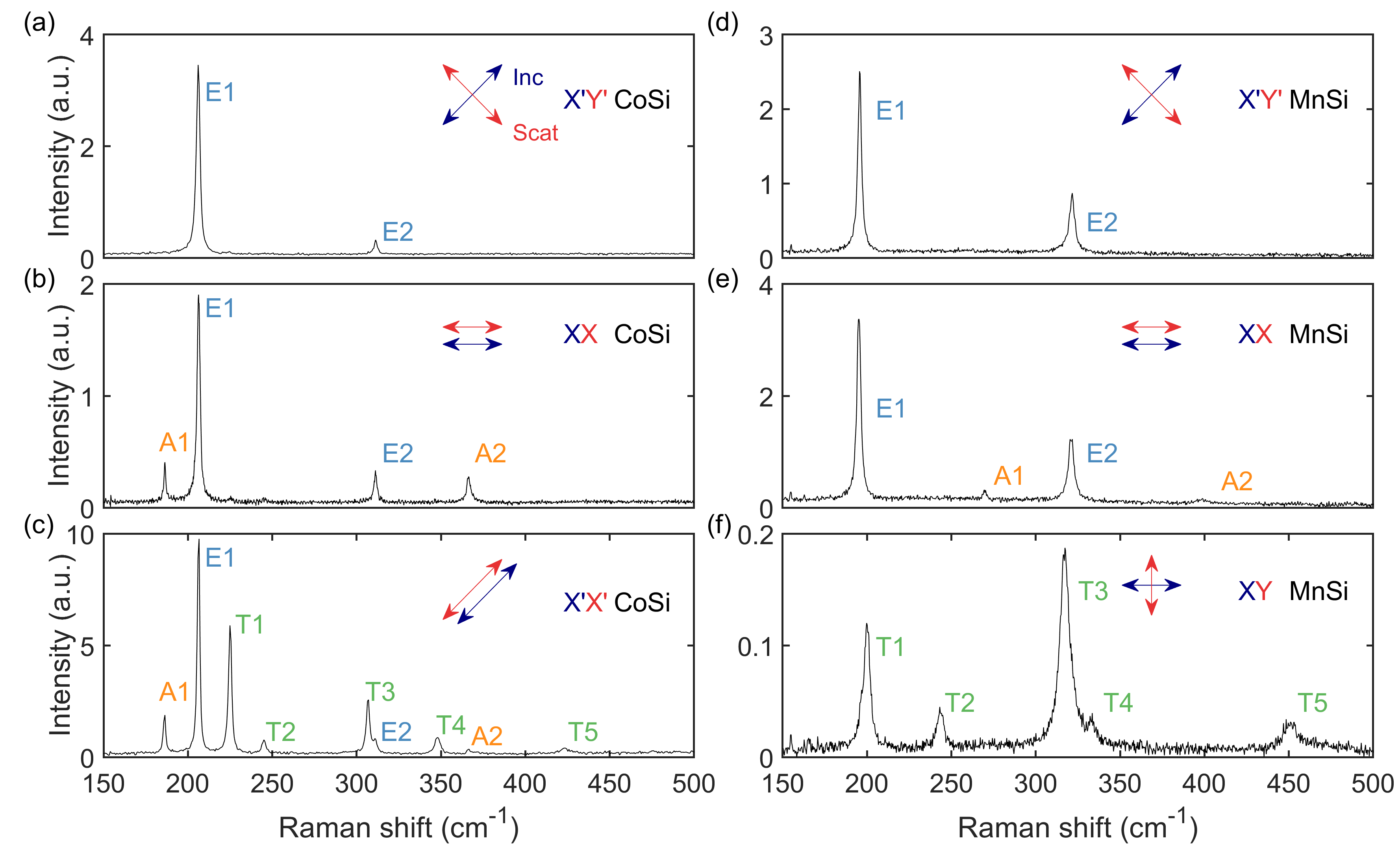}
    \caption{\label{figs3}(a)--(f) Raman spectra of the optical phonons in CoSi and MnSi. The $A,E,T$ irreps at $\Gamma$ are identified by utilizing different polarizations, so as to double-check the degeneracy of each band-crossing. X'Y' directions are in 45${}^{\circ}$ angle with XY.}
  \end{figure}
  \begin{figure}[!ht]
    \includegraphics[width=0.65\textwidth]{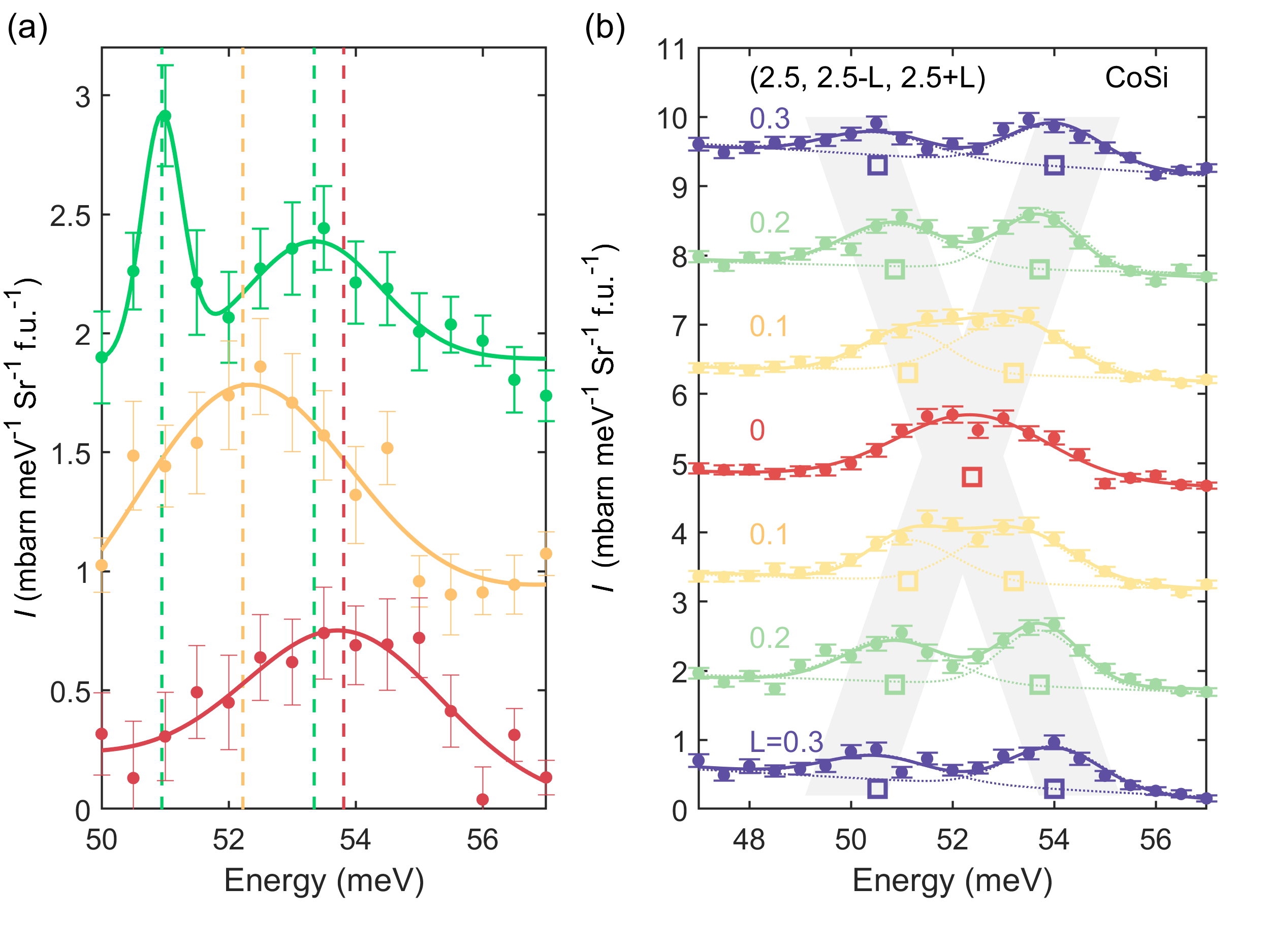}
    \caption{\label{figs4}{(a) Gaussian fits using the same data in Fig. 2(a) in the main text, assuming two/one peak for the green/red data points. The fits yield three different peak positions illustrated by the colored dash lines. (b) Energy cuts of the INS data at a series of $\mathbf{Q}$ points in Fig. 2(c) of main text. The open squares denote the peak positions obtained from two-peak fits, under the constraint that equivalent $\mathbf{q}$ positions must have same energies and peak widths. As indicated by the blue cross, the peak positions are further fitted collectively into an X-shape linear dispersion near the R point, in consistance with a charge-2 Dirac point. Data in (a) and (b) are offset for clarity.}}
  \end{figure}
\clearpage
\section{The fitted-force-constant model based on DFPT}
  First of all, the force constant matrices in MnSi and CoSi are calculated by DFPT with \textsc{vasp}. From the force constants, phonon dispersions and eigenmodes can be obtained by solving a classical, harmonic ``spring-ball'' model. Although capturing most features in the phonon dispersions qualitatively, the \textit{ab initio} results are not precise enough in predicting certain energy and intensity values.

\begin{figure}[!ht]
    \includegraphics[width=0.85\textwidth]{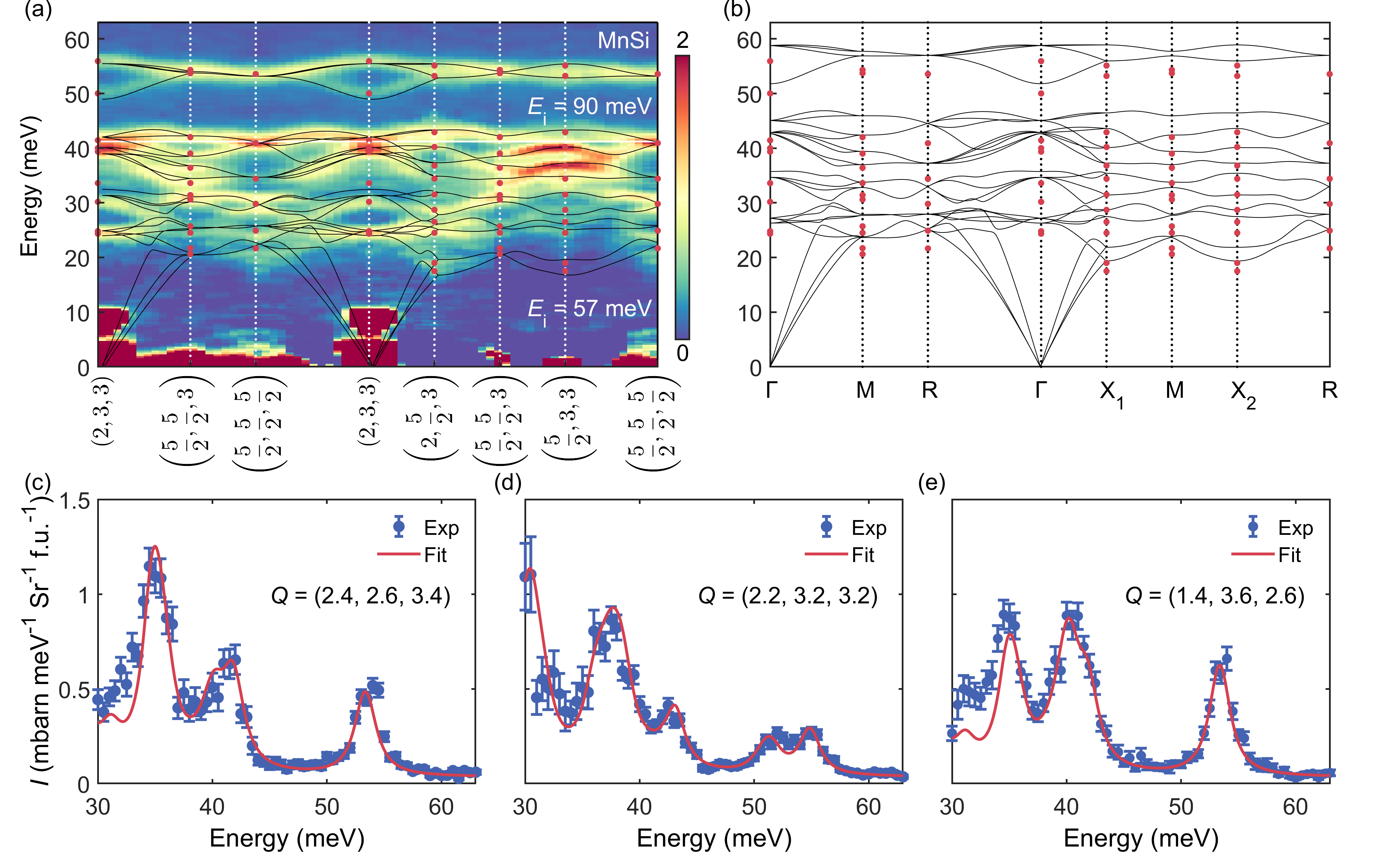}
    \caption{\label{figs5}(a) INS spectra of phonons in MnSi. Red dots show the energy values at $\Gamma$, X, M, R. The black lines illustrate phonon dispersions simulated with the fitted-force-constant model, in good agreement with INS results. (b) Unadjusted phonon dispersions from DFPT outputs, with significant difference with experiments. (c)--(e) Measured and calculated INS intensities at selected $\mathbf{Q}$-points, showing remarkable accordance.}
    \vspace{0.2cm}
  \end{figure}

  To solve this, we first extract the energy values of all phonon branches at high symmetry points $\Gamma$, X, M, R in the BZ, by visually inspecting INS spectra and taking average over many BZs. The results are listed in Table \ref{tab1}.
  Retrieved phonon energies of MnSi are also shown by red dots in Fig.~\ref{figs5}(a) and (b), together with the intensity spectra. The original phonon dispersions calculated directly from \textit{ab initio} outputs are displayed in Fig.~\ref{figs5}(b). It is clear that phonon frequencies are overestimated globally, and $\sim 10\%$ variations are possible at some points in the BZ. The errors still exist partially after performing a global rescale in energy, which accounts for the inaccuracy in relaxed lattice constants.
  \newcolumntype{Y}{>{\centering\arraybackslash}X}
  \begin{table}[b]
  \vspace{0.5cm}
  \begin{tabularx}{1\textwidth}{Y|Y|YYYYYYYYYYYY}
  \toprule
  \multirow{4}{*}{MnSi} & $\Gamma$ & \textcolor{white}{0.}0\textsuperscript{T}\textcolor{white}{0} & 24.3\textsuperscript{E} & 24.8\textsuperscript{T} & 30.2\textsuperscript{T} & 33.6\textsuperscript{A} & 39.4\textsuperscript{T} & 40.0\textsuperscript{E} & 41.4\textsuperscript{T} & 49.7\textsuperscript{A} & 55.9\textsuperscript{T} & \textcolor{white}{55.9\textsuperscript{T}} & \textcolor{white}{55.9\textsuperscript{T}} \\\cline{2-14}
  & $\rm X$ & 17.5 & 19.0    & 24.5    & 26.5    & 28.7    & 31.5    & 34.4    & 36.8    & 40.2    & 42.9    & 53.2 & 55.1 \\\cline{2-14}
  & $\rm M$ & 20.6 & 21.7    & 24.5    & 25.7    & 30.6    & 31.4    & 33.6    & 36.4    & 39.0    & 42.0    & 53.7 & 54.3 \\\cline{2-14}
  & $\rm R$ & 21.7 & 24.9    & 29.8    & 34.4    & 40.9    & 53.6    &         &         &         &         &      &      \\\midrule
  \multirow{4}{*}{CoSi} & $\Gamma$ & 0\textsuperscript{T} & 23.2\textsuperscript{A} & 25.7\textsuperscript{E} & 28.0\textsuperscript{T} & 30.5\textsuperscript{T} & 38.2\textsuperscript{T} & 38.7\textsuperscript{E} & 43.3\textsuperscript{T} & 45.5\textsuperscript{A} & 52.5\textsuperscript{T} &      &      \\\cline{2-14}
  & $\rm X$ & 17.0 & 18.5    & 23.2    & 26.3    & 29.7    & 31.3    & 35.5    & 37.6    & 40.9    & 45.7    & 49.2 & 54.2 \\\cline{2-14}
  & $\rm M$ & 18.0 & 20.3    & 22.9    & 25.5    & 29.5    & 32.1    & 33.9    & 36.4    & 40.6    & 45.9    & 50.3 & 53.7 \\\cline{2-14}
  & $\rm R$ & 20.7 & 22.8    & 31.5    & 33.5    & 42.8    & 52.7    &         &         &         &         &      & \\
  \bottomrule
  \end{tabularx}
  \caption{\label{tab1}Phonon energies (in meV) of all bands at high symmetry points ($\Gamma$, X, M, R), obtained from experimental INS spectra. The superscripts $A, E, T$ stand for 1, 2, 3-fold degeneracy at $\Gamma$, determined by Raman spectroscopy. Errors at all points are roughly $0.5$ meV due to the limitation of resolution.}
  \vspace{0.5cm}
  \end{table}

  We found that such deviations can be greatly reduced by adjusting force constants and performing parametric fits. First, the force constant matrices between atom $d$ and $d^{\prime}$, $\mathbb{K}_{jj^{\prime}dd^{\prime}}$ are categorized into equivalent classes according to space group symmetry, and arranged by their matrix norm $|\mathbb{K}|$ (``strength of springs''). Then, each of the dominating $N$ classes are multiplied by an adjustable variable $a_i$, while all other insignificant force constants are set to zero. In this way, the dynamical matrix $\mathbb{D}(\mathbf{q})$ is constructed by
    \begin{align}
      \mathbb{D}(\mathbf{q})=a_1\cdot\sum (\mathbb{K}\text{ in class \#1})+a_2\cdot\sum (\text{class \#2})+\cdots +a_N\cdot\sum (\text{class \#N})
    \end{align}
  and diagonalized to get the phonon dispersions. Finally, a $\chi^2$-fit is performed on $\{a_i\}$ to minimize the difference between calculated and experimental energy datasets. The $\chi^2$ is defined as
    \begin{align}
      \chi^2=\sum_{\{ E_i\}}\left(\frac{E_i^{\text{calc}}-E_i^{\text{exp}}}{\Delta E_i}\right)^2
    \end{align}
  and minimized by the conjugate descend method. All $\Delta E_i$ are assumed to be $0.5$ meV. We also reassured the algorithm's convergence by testing with different initial values.

  Detailed information of the parametric fits can be found in Table \ref{tab2}. It is surprising that only very few independent matrices are needed ($N=9$ for MnSi and $N=11$ for CoSi) to greatly reduce $\chi^2$ and reproduce experimental data to a remarkable extent, either globally [Fig.\ref{figs5}(a)] or at certain $\mathbf{Q}$-points [Fig.\ref{figs5}(c)-(e)]. Since we set all other force constants to zero, we are actually trying to find out which types of ``springs'' are most important in real crystals. Generally, atom pairs with greater distances have smaller force constants, i.e. weaker interactions; but it is not always the case. Ranking by their optimized strengths, fitting parameters $\{a_i\}$ and atom distances of the leading force constants are listed together in Table \ref{tab2}.

  \begin{table}[t]
    \begin{tabularx}{1.\textwidth}{YYY|YYY|YYY}
    \toprule
    Material & $N$ & $\chi^2$ & \multicolumn{6}{c}{Optimized bond strengths $|\mathbb{K}|_i$ \textcolor{blue}{(fitting parameters $\{a_i\}$)}}\\\midrule
    MnSi & 0 & 1462 & \multicolumn{6}{c}{-------- (DFPT output)}\\
    MnSi & 1 & 112.2 & \multicolumn{6}{c}{\textcolor{blue}{0.857} (global rescale)}\\\hline
    \multirow{5}{*}{\textbf{MnSi}} & \multirow{5}{*}{\textbf{9}} & \multirow{5}{*}{\textbf{56.9}} & 17.61 \textcolor{blue}{(0.896)} & Si--Si & 0.00 \AA & 1.11 \textcolor{blue}{(0.719)} & Mn--Mn & 2.79 \AA\\
    & & & 14.60 \textcolor{blue}{(0.810)} & Mn--Mn & 0.00 \AA & 1.10 \textcolor{blue}{(0.782)} & Si--Si & 2.82 \AA\\
    & & & 6.93 \textcolor{blue}{(0.949)} & Mn--Si & 2.31 \AA & 0.79 \textcolor{blue}{(1.358)} & Mn--Mn & 4.14 \AA\\
    & & & 4.25 \textcolor{blue}{(0.796)} & Mn--Si & 2.40 \AA & 0.51 \textcolor{blue}{(1.387)} & Mn--Mn & 4.57 \AA\\
    & & & 2.12 \textcolor{blue}{(0.956)} & Mn--Si & 2.54 \AA & & \\\midrule
    CoSi & 0 & 465.6 & \multicolumn{6}{c}{-------- (DFPT output)}\\
    CoSi & 1 & 379.7 & \multicolumn{6}{c}{\textcolor{blue}{0.960} (global rescale)}\\\hline
    \multirow{6}{*}{\textbf{CoSi}} & \multirow{6}{*}{\textbf{11}} & \multirow{6}{*}{\textbf{70.6}} & 16.17 \textcolor{blue}{(0.989)} & Si--Si & 0.00 \AA & 1.37 \textcolor{blue}{(1.011)} & Co--Si & 2.44 \AA\\
    & & & 12.88 \textcolor{blue}{(0.909)} & Co--Co & 0.00 \AA & 0.88 \textcolor{blue}{(1.323)} & Co--Co & 3.97 \AA\\
    & & & 6.12 \textcolor{blue}{(1.477)} & Co--Si & 2.33 \AA & 0.62 \textcolor{blue}{(0.625)} & Si--Si & 3.87 \AA\\
    & & & 3.54 \textcolor{blue}{(1.038)} & Co--Si & 2.34 \AA & 0.52 \textcolor{blue}{(1.235)} & Si--Si & 4.47 \AA\\
    & & & 1.95 \textcolor{blue}{(0.815)} & Co--Co & 2.74 \AA & 0.35 \textcolor{blue}{(0.695)} & Si--Si & 4.45 \AA\\
    & & & 1.88 \textcolor{blue}{(1.057)} & Si--Si & 2.75 \AA & & \\
    \bottomrule
    \end{tabularx}
    \caption{\label{tab2}Results of the $\chi^2$-fits and the optimized parameter sets for MnSi and CoSi. The optimized strengths (norms of force constant matrice $|\mathbb{K}|_i$, in eV/\AA$^2$) of different types of bonds are listed in a descending order. The fitting parameters $\{a_i\}$ are shown in blue and in parentheses. Corresponding bond types and bond lengths are also listed.}
    \vspace{0.2cm}
  \end{table}

\newpage
\section{Determination of Chern numbers from inelastic neutron scattering intensity}
\subsection{Generic two-fold Weyl points}
  Aside from some unimportant constant coefficients, the coherent dynamical structure factor  of the $i$\textsuperscript{th} phonon mode, $S_{\text{coh}}^{(i)}(\mathbf{Q},\omega)$, is dominated by the polarization vector $\bm{\xi}^{(i)}_d(\mathbf{q})$\cite{Lovesey1984,squires_2012}.
  \begin{align}
  S_{\text{coh}}^{(i)}(\mathbf{Q},\omega)\propto |F^{(i)}_{\text{coh}}(\mathbf{Q})|^2= \left|\sum_d\frac{b_{d,\text{coh}}}{\sqrt{m_d}}\mathbf{Q}\cdot\bm{\xi}^{(i)}_d(\mathbf{q}) e^{i\mathbf{Q}\cdot \mathbf{r}_d}\right|^2.
  \end{align}
  In this formula, $\mathbf{Q}=\mathbf{q}+\mathbf{G}$ is the total momentum transfer, and $m_d, \mathbf{r}_d, b_{d,\text{coh}}$ denote the mass, position and coherent scattering length \cite{shirane_shapiro_tranquada_2002} of the $d$\textsuperscript{th} atom in a single unit cell. The most important term $\mathbf{Q}\cdot\bm{\xi}^{(i)}_d(\mathbf{q})$ is decided by the interaction between neutrons and point-like nuclei.

  For an arbitrary two-fold band crossing at $\mathbf{q}=0$, we assume that the two phonon eigenmodes at $\mathbf{q}=0$ are $\bm{\xi}^{(i=1,2)}$. Under this basis, the effective Hamiltonian of a two-fold Weyl point is denoted by a $2\times 2$ Hermitian matrix $H_{2\times2}(\mathbf{q})=\mathbf{f}(\mathbf{q})\cdot \bm{\sigma}+f_0(\mathbf{q})\sigma_0$. Then, eigenvector of the upper branch  at finite $\mathbf{q}$ is
  \begin{align}
    \bm{\xi}^{\text{up}}(\mathbf{q})=\psi_1(\mathbf{q})\bm{\xi}^{(1)}+\psi_2(\mathbf{q})\bm{\xi}^{(2)}.
  \end{align}
  In the close vicinity of the Weyl point, we consider the limit $|\mathbf{q}|\ll |\mathbf{G}|$ and $|\mathbf{q}|\ll 2\pi/a$, so that the upper band intensity is approximately
  \begin{align}
    \nonumber
  S_{\text{coh}}^{\text{up}}(\mathbf{Q},\omega)&\propto \left|\sum_d\frac{b_{d,\text{coh}}}{\sqrt{m_d}}\mathbf{G}\cdot\bm{\xi}^{\text{up}}_d(\mathbf{q}) e^{i\mathbf{G}\cdot \mathbf{r}_d}\right|^2\\
  &=\left|\psi_1(\mathbf{q}) W^{(1)}(\mathbf{G})+\psi_2(\mathbf{q}) W^{(2)}(\mathbf{G})\right|^2,
  \end{align}
  where $W^{(i=1,2)}$ are defined as
  \begin{align}
  W^{(i)}(\mathbf{G})\equiv\sum_{d}\frac{b_{d,\text{coh}}}{\sqrt{m_d}}\mathbf{G}\cdot\bm{\xi}^{(i)}_d e^{i\mathbf{G}\cdot \mathbf{r}_d}.
  \end{align}
  From the expressions above, it is clear that $S_{\text{coh}}^{\text{up}}(\mathbf{Q},\omega)$ is nothing but the inner product of two complex vectors: spinor  $\bm{\xi}^{\text{up}}(\mathbf{q})=\left(\psi_1(\mathbf{q}),\psi_2(\mathbf{q})\right)^{\text{T}}$ and $\mathbf{W}(\mathbf{G})=\left(W^{(1)}(\mathbf{G})^*,W^{(2)}(\mathbf{G})^*\right)^{\text{T}}$, a vector independent of $\mathbf{q}$ but varying among different BZs. From this formula, the INS intensity would reach maximum when the two vectors are parallel, i.e.
  \begin{align}
    \left(\psi_1(\mathbf{q}),\psi_2(\mathbf{q})\right)=c\left(W^{(1)}(\mathbf{G})^*,W^{(2)}(\mathbf{G})^*\right)
  \end{align}
  and reach zero when they are perpendicular, i.e.
  \begin{align}
    \left(\psi_1(\mathbf{q}),\psi_2(\mathbf{q})\right)=c\left(-W^{(2)}(\mathbf{G}),W^{(1)}(\mathbf{G})\right).
  \end{align}

  Following the definition of pseudospin operators in the main text, the spinor  $\bm{\xi}^{\text{up}}(\mathbf{q})$ is represented by a unit 3-dimensional real vector $\mathbf{S}(\mathbf{q})$ on the Bloch sphere
\begin{align}
S_{i}(\mathbf{q}) \equiv \frac{f_{i}(\mathbf{q})}{|\mathbf{f}(\mathbf{q})|}=\left\langle\xi^{\text {up}}\left|\sigma_{i}\right|\xi^{\text {up}}\right\rangle.
\end{align}
Similarly, the complex vector $\mathbf{W}(\mathbf{G})$ can also be represented by a 3-dimensional real vector $\mathbf{V}(\mathbf{G})$ 
\begin{align}
V_{i}(\mathbf{G}) =\left\langle\mathbf{W}(\mathbf{G})\left|\sigma_{i}\right|\mathbf{W}(\mathbf{G})\right\rangle,
\end{align}
whose modulus $|\mathbf{V}|$ is not necessarily unitary. In this notation, the scattering intensity is proportional to the inner product
  \begin{align}
    \nonumber
    S_{\text{coh}}^{\text{up}}(\mathbf{Q},\omega)&\propto\left|\psi_1(\mathbf{q}) W^{(1)}(\mathbf{G})+\psi_2(\mathbf{q}) W^{(2)}(\mathbf{G})\right|^2\\\nonumber
    &=\frac{1}{2}(|\mathbf{V}|+\mathbf{S}(\mathbf{q})\cdot \mathbf{V}(\mathbf{G}))\\
    &\propto|\mathbf{V}| \left(1+\frac{\mathbf{S}(\mathbf{q})\cdot \mathbf{V}(\mathbf{G})}{|\mathbf{S}(\mathbf{q})||\mathbf{V}(\mathbf{G})|}\right).
  \end{align}
  Now we arrive at the formula described in the main text.

  Next, we consider a closed surface $\mathbb{S}_{\mathbf{q}}$ in $\mathbf{q}$-space that encloses the origin. For two-fold Weyl points, the upper (lower) band is associated with non-zero Chern number $\pm C$, defined by the total Berry flux going in or out of the surface. If momentum $\mathbf{q}$ runs over the surface, the pseudospin-1/2 spinor $\left(\psi_1(\mathbf{q}),\psi_2(\mathbf{q})\right)^T$ will wrap the $\mathbb{CP}^1$-space exactly $|C|$ times, or equivalently, the pseudospin $\mathbf{S}$ will wrap the Bloch sphere $\mathbb{S}_{\mathrm{B}}$ exactly $|C|$ times. Therefore, we conclude that there should be at least $|C|$ momenta on the $\mathbf{q}$-sphere where intensity reaches maximum, and at least $|C|$ momenta where intensity reaches zero. As mentioned in the main text, the approximation of $\mathbf{Q}=\mathbf{q}+\mathbf{G}\approx \mathbf{G}$ can also be released since $\mathbf{Q}$ has trivial topology on the surface $\mathbb{S}_{\mathbf{q}}$ as long as the origin of $\mathbf{Q}$ is not enclosed.

\subsection{Charge-2 Dirac points}
  We next generalize this deduction to the Charge-2 Dirac points at R. Although being four-fold degenerated, the Hamiltonian
  $H_4(\mathbf{q})\propto\left(\begin{smallmatrix}
    \mathbf{q}\cdot \bm{\sigma} & 0\\
    0 & \mathbf{q}\cdot \bm{\sigma}
  \end{smallmatrix}\right)$
  is the direct sum of two identical spin-1/2 Weyl points, each with Chern number $C=\pm 1$ \cite{Zhangtiantian2018}. So eigenvectors of the two upper branches are
  \begin{align}\begin{split}
    \bm{\xi}^{\text{up}}(\mathbf{q})&=\psi_1(\mathbf{q})\bm{\xi}^{(1)}+\psi_2(\mathbf{q})\bm{\xi}^{(2)},\\
    \bm{\xi}^{\text{up}\prime}(\mathbf{q})&=\psi_1(\mathbf{q})\bm{\xi}^{(3)}+\psi_2(\mathbf{q})\bm{\xi}^{(4)}.
    \end{split}
  \end{align}
  Again we assume $|\mathbf{q}|\ll |\mathbf{G}|$ and $|\mathbf{q}|\ll 2\pi/a$, and the total intensity of the two upper bands is
    \begin{align}
      \nonumber
    S_{\text{coh}}^{\text{up}}(\mathbf{Q},\omega)&\propto\left|\psi_1(\mathbf{q}) V^{(1)}(\mathbf{G})+\psi_2(\mathbf{q}) V^{(2)}(\mathbf{G})\right|^2+\left|\psi_1(\mathbf{q}) V^{(3)}(\mathbf{G})+\psi_2(\mathbf{q}) V^{(4)}(\mathbf{G})\right|^2\\
    &=\left(\psi_1(\mathbf{q}),\psi_2(\mathbf{q})\right)^*\cdot \mathbb{V}(\mathbf{G})\cdot\begin{pmatrix}
      \psi_1(\mathbf{q})\\
      \psi_2(\mathbf{q})
    \end{pmatrix},
    \end{align}
    where $\mathbb{V}(\mathbf{G})$ is a $2\times2$ Hermitian matrix
\begin{align}
\begin{pmatrix}
| V^{(1)}|^2+| V^{(3)}|^2 & V^{(1)} V^{(2)*}+ V^{(3)} V^{(4)*}\\
V^{(1)*} V^{(2)}+ V^{(3)*} V^{(4)} & | V^{(2)}|^2 + | V^{(4)}|^2
\end{pmatrix}
\end{align}
It can be written as $\mathbb{V}(\mathbf{G})=\mathbf{V}(\mathbf{G})\cdot\bm{\sigma}+V_0$, where $V_0$ is equal to or larger than $|\mathbf{V}|$ because $\mathbb{V}(\mathbf{G})$ is semi positive definite. Again, it is also represented by a 3-dimensional real vector $\mathbf{V}(\mathbf{G})$, but with an extra constant $V_0$. In this case the scattering intensity would be
    \begin{align}
      \nonumber
      S_{\text{coh}}^{\text{up}}(\mathbf{Q},\omega)&\propto 2V_0+|\mathbf{V}|+\mathbf{S}(\mathbf{q})\cdot\mathbf{V}(\mathbf{G})\\
      &=2V_0+|\mathbf{V}|\left(1+\frac{\mathbf{S}(\mathbf{q})\cdot \mathbf{V}(\mathbf{G})}{|\mathbf{S}(\mathbf{q})||\mathbf{V}(\mathbf{G})|}\right).
    \end{align}
    where $V_0$ is a non-negative constant.

    In the case of charge-2 Dirac points, each spin-1/2 copy in the Hamiltonian has Chern number $\pm 1$. Similar to the arguments above, the pseudospin $\mathbf{S}$ will wrap the Bloch sphere exactly $|C|$ times on a $\mathbf{q}$-surface that encloses the origin. Therefore, the total INS intensity of the two upper bands will have exactly one maximum and two minimum on the $\mathbf{q}$-sphere. The only difference due to the two spin-1/2 components is that the smallest intensity may be larger than zero.
\end{document}